\documentclass[preprintnumbers,showpacs,amsmath,amssymb,floatfix,prd,onecolumn,superscriptaddress,nofootinbib]{revtex4}
\usepackage{bbm}
\usepackage{graphicx}
\usepackage{epsfig}
\usepackage{bm}
\usepackage{amsfonts}
\usepackage{epstopdf}
\begin{document}
\title{Fermionic Casimir effect with helix boundary condition}
\author{Xiang-hua Zhai}\email{zhaixh@shnu.edu.cn}

\author{Xin-zhou Li}\email{kychz@shnu.edu.cn}

\author{Chao-Jun Feng}
\email{fengcj@shnu.edu.cn}

\affiliation{Shanghai United Center for Astrophysics (SUCA), \\
Shanghai Normal University,
    100 Guilin Road, Shanghai 200234, China}

\begin{abstract}
 In this paper, we consider the fermionic Casimir effect
under a new type of space-time topology using the concept of
quotient topology. The relation between the new topology and that in
Ref. \cite{Feng,Zhai3} is something like that between a M\"obius
strip and a cylindric. We obtain the exact results of the Casimir
energy and force for the massless and massive Dirac fields in the
($D+1$)-dimensional space-time. For both massless and massive cases,
there is a $Z_2$ symmetry for the Casimir energy. To see the effect
of the mass, we compare the result with that of the massless one and
we found that the Casimir force approaches the result of the force
in the massless case when the mass tends to zero and vanishes when
the mass tends to infinity.
\end{abstract}
\maketitle
\section{Introduction}\label{sec:1}
Casimir's computation of the force between two neutral, parallel
conducting plates\cite{Casimir} originally inspired much theoretical
interest as macroscopic manifestation of quantum fluctuation of the
field in vacuum. However, the Casimir force arises not only in the
presence of material boundaries, but also in spaces with
non-Euclidean topology\cite{Bordag} and the attractive or repulsive
nature of the Casimir force may depend on the topology of spacetime.
The simplest example of the Casimir effect of topological origin is
the scalar field on a flat manifold with topology of a circle
$\textit{S}^1$. The topology of $\textit{S}^1$ causes the
periodicity condition $\phi(t,0)=\phi(t,C)$ for a Hermitian scale
field $\phi(t,x)$, where $C$ is the circumference of $\textit{S}^1$,
imposed on the wave function which is of the same kind as those due
to boundary and resulting in an attractive Casimir force. Similarly,
the antiperiodic conditions can be drawn on a M\"{o}bius strip and
bring about the repulsive Casimir force as a result. Recently, the
topology of the helix boundary conditions is investigated in
Ref.\cite{Feng}. We find that the Casimir force is very much like
the force on a spring that obeys the Hooke's law in mechanics.
However, in this case, the force comes from a quantum effect, so we
would like to call this structure a quantum spring. The force is
attractive in both massless and massive scalar cases for this
structure\cite{Zhai3}.

It is worth noting that the concept of quotient topology is very
useful for concrete application. We consider a surjective mapping
$f$ from a topological space $X$ onto a set $Y$. The quotient
topology on $Y$ with respect to $f$ is given in \cite{Munkres}.
Surjective mapping can be easily obtained when we use the
equivalence classes of some equivalence relation $\sim$. Thus, we
let $X/\sim$ denote the set of equivalence classes and define $f:
X\rightarrow X/\sim$ by $f(x)=[x]$ the equivalence class containing
$x$. $X/\sim$ with the quotient topology is called to be obtained
from $X$ by topological identification. For example, if we take the
unit square $I^2=\{(x_1, x_2); 0\leq x_1,x_2 \leq 1\}$ in
$\mathbb{R}^2$ with the induced topology and define an
equivalence relation $\sim$ on $I^2$ by
\begin{eqnarray}
& &(x_1,x_2)\sim(x_1^{\prime}, x_2^{\prime})\Leftrightarrow
(x_1,x_2)=(x_1^{\prime}, x_2^{\prime}) \nonumber\\
&\mathrm{or}& \{x_1, x_1^{\prime}\}=\{0,1\}\hspace{0.3cm} \mathrm{and}\hspace{0.3cm}
x_2=x_2^{\prime}
\end{eqnarray}
then $I^2/\sim$ with the quotient topology is homomorphic to the
cylinder $C$
\begin{equation}
C=\{(x_1,x_2,x_3)\in \mathbb{R}^3; x_1^2+x_2^2=1, |x_3|\leq 1\}
\end{equation}
The boundary condition $\phi(t,0,x_2)=\phi(t,1,x_2)$ can be drawn on
a cylinder.

The $\zeta$-function regularization procedure is a powerful and
elegant technique for the Casimir effect. Rigorous extension of the
proof of Epstein $\zeta$-function regularization has been discussed
in \cite{Elizalde}. Vacuum polarization in the background of on
string was first considered in \cite{Helliwell:1986hs}. The
generalized $\zeta$-function has many interesting applications,
e.g., in the piecewise string \cite{Brevik0,Li:1990bz,Brevik}.
Similar analysis has been applied to cosmology entropy
\cite{Brevik1}, p-branes \cite{Shi:1991qc}, rectangular cavity
\cite{Li,Li2}, or pistons
\cite{Cavalcanti,Hertzberg,Zhai,Zhai2,Lim}. Casimir effect for a
fermionic field is of interest in considering, for example, the
structure of proton in particle physics \cite{Chodos,Elizalde2}
while the thermofield dynamics of Casimir effect for Dirac fermion
fields has been studied \cite{Queiroz}  and recently for Majorana
fermion fields \cite{Erdas}. The Casimir piston of fermion is also
studied \cite{Oikonomou}. And the fermionic Casimir effect in the
presence of compact dimensions has been recently considered in Refs.
\cite{Bellucci} and \cite{Elizalde3}.

In this paper, we consider the fermionic Casimir effect under a new
type of space-time topology using the concept of quotient topology.
The relation between the new helix topology and that we use in
\cite{Feng,Zhai3} is something like that between a M\"{o}bius strip
and a cylindric. We obtain the exact results of the Casimir energy
and Casimir force for the massless and massive Dirac fields in the
($D+1$)-dimensional space-time.

\section{The vacuum energy density for a fermionic field }

As mentioned in Sec. 1, the Casimir effect arises not only in
presence of material boundaries, but also in space with nontrivial
topology. We consider topological space $X$ as follows
\begin{equation}
X=\bigcup_{\mathbbm{u}\in \mathrm{\Lambda}''}\{C_0+\mathbbm{u}\}
\end{equation}
in $\mathcal{M}^{D+1}$ with the induced topology and define an
equivalence relation $\sim$ on $X$ by
\begin{equation}
(x^1,x^2)\sim(x^1-2a, x^2+2h)
\end{equation}
then $X/\sim$ with the quotient topology is homomorphic to helix
topology. Here, $\mathrm{\Lambda}''$ and unit cylinder-cell $C_0$
\cite{Feng,Zhai3} are
\begin{equation}\label{sub2}
  \mathrm{ \Lambda}'' = \left\{ ~ n(\mathbbm{e}_2 - \mathbbm{e}_1) ~|~ n \in \mathcal{Z} ~\right\} \,.
\end{equation}
and
\begin{eqnarray}
   C_0 &=& \bigg\{ \sum_{i=0}^{D}x^i \mathbbm{e}_i ~|~ 0\leq x^1 < a,
 -h\leq x^2 < 0 ,\nonumber\\
 &-&\infty <x^0<\infty, -\frac{L}{2} \leq x^T\leq \frac{L}{2}\bigg\} \,,\label{cell}
\end{eqnarray}
where $T = 3,\cdots, D$. Next, we discuss what are referred to as
anti-helix conditions imposed on a field $\psi$,
\begin{equation}
\psi(t, x^1+a,x^2,x^T)=-\psi(t,x^1,x^2+h,x^T),
\end{equation}
where the field returns to the same value after traveling distances
$2a$ at the $x^1$-direction and $2h$ at the $x^2$-direction. It is
notable that a spinor wave function is anti-helix and takes its
initial value after traveling distances $2a$ and $2h$ respectively.
In other words, the anti-helix conditions are imposed on the field,
which returns to the same field value $\psi(t,
x^1+2a,x^2,x^T)=\psi(t,x^1,x^2+2h,x^T)$ only after two round trips.
Therefore, the boundary condition (7) can be induced by $X/\sim$
with the quotient topology.

In calculations on the Casimir effect, extensive use is made of
eigenfunctions and eigenvalues of the corresponding field equation.
A spin-1/2 field $\psi(t, x^\alpha, x^T)$ defined in the
($D+1$)-dimensional flat space-time satisfies the Dirac equation:
\begin{equation}\label{eom}
    i\gamma^{\mu}\partial_{\mu}\psi - m_0\psi = 0 \,,
\end{equation}
where $\alpha=1,2; T=3,\cdots, D$; $\mu=(t,\alpha,T)$ and $m_0$ is
the mass of the Dirac field. $\gamma^{\mu}$ are $N\times N$ Dirac
matrices with $N=2^{[(D+1)/2]}$ where the square brackets mean the
integer part of the enclosed expression. We will assume that these
matrices are given in the chiral representation:

\begin{equation}
\gamma^0=\Bigg(\begin{array}{clrr}
      1 &\hspace{0.2cm} 0  \\       0  & -1  \end{array}\Bigg),\gamma^k=\Bigg(\begin{array}{clrr}
      0 & \sigma_k  \\       -\sigma_k^{+}  & 0  \end{array}\Bigg), k=1,2,\cdots, D
\end{equation}
with the relation
$\sigma_{\mu}\sigma_{\nu}^{+}+\sigma_{\nu}\sigma_{\mu}^{+}=2\delta_{\mu\nu}$.
Under the boundary condition (7), the solutions of the field can be
presented as
\begin{equation}
\psi^{(+)}=\mathcal{N}^{(+)}e^{-i\omega t}\Bigg(\begin{array}{clrr}
      e^{i(k_x x+k_z z+k_T x^T)}\varphi_{(\alpha)}  \\    -i\mbox{\boldmath$\sigma$}^{+}\cdot\mbox{\boldmath$\nabla$} e^{i(k_x x+k_z z+k_T x^T)}\varphi_{(\alpha)} /(\omega+m_0)   \end{array}\Bigg),
\end{equation}
and
\begin{equation}
\psi^{(-)}=\mathcal{N}^{(-)}e^{i\omega t}\Bigg(\begin{array}{clrr}
       i\mbox{\boldmath$\sigma$}\cdot \mbox{\boldmath$\nabla$}e^{i(k_x x+k_z z+k_T x^T)}\chi_{(\alpha)} /(\omega+m_0) \\  e^{i(k_x x+k_z z+k_T x^T)}\chi_{(\alpha)}     \end{array}\Bigg),
\end{equation}
where $\mbox{\boldmath$\sigma$}=(\sigma_1,\cdots, \sigma_D), x^1=x,
x^2=z$ and $\mathcal{N}^{(\pm)}$ is a normalization factor, and
$\varphi_{(\alpha)}, \chi_{(\alpha)}$ are one-column constant matrices
having $2^{[(D+1)/2]}-1$ rows with the element
$\delta_{\alpha\beta}, \alpha,\beta=1,\cdots, 2^{[(D+1)/2]}-1$.

From Eqs. (8)-(11), we have
\begin{eqnarray}\label{energy}
    \omega_n^2 &=& k_{T}^2 + k_x^2 + \left( -\frac{2\pi (n+\frac 12)}{h}+\frac{k_x}{h}a \right)^2+m_0^2 \nonumber\\
    &=& k_{T}^2 + k_z^2 + \left( \frac{2\pi (n+\frac 12)}{a}+\frac{k_z}{a}h
    \right)^2+m_0^2 \,.
\end{eqnarray}
Here, $k_x$ and $k_z$ satisfy
\begin{equation}\label{kxkz}
    a k_x - hk_z = 2\left (n+\frac 12\right )\pi\,, (n=0,\pm1,\pm2,\cdots) \,.
\end{equation}
In the ground state (vacuum), each of these modes contributes an
energy of $\omega_n/2$. The energy density of the field in
($D+1$)-dimensional space-time is thus given by

\begin{eqnarray}
\nonumber
  E^{D}=&-&\frac{N}{2 a}
  \int \frac{d^{D-1}k}{(2\pi)^{D-1}} \\
  &\times&\sum_{n=-\infty}^{\infty} \sqrt{k_T^2 + k_z^2 + \left( \frac{2\pi (n+\frac 12)}{a}
  +\frac{k_z}{a}h \right)^2 +m_0^2 }\nonumber\\& & \label{tot energy}
\end{eqnarray}

\noindent where we have assumed $a\neq 0$ without losing
generalities.

Eq.(\ref{tot energy}) can be rewritten as
\begin{eqnarray}
\nonumber
  E^{D}=&-& \frac{N}{2 a \sqrt{\gamma}}
  \int \frac{d^{D-1}u}{(2\pi)^{D-1}} \\
  &\times&\sum_{n=-\infty}^{\infty} \sqrt{u^2 + \left( \frac{2\pi (n+\frac 12)}
  {a \sqrt{\gamma}}\right)^2+m_0^2  }\nonumber \,, \\&&\label{re-tot energy}
\end{eqnarray}

\noindent where
\begin{equation}\label{gamma}
\gamma \equiv 1+ \frac {h^2}{a^2}.
\end{equation}

\noindent Using the mathematical identity,
\begin{equation}
\int_{-\infty}^{\infty} f(u) d^{D-1} u=\frac {2\pi^{\frac
{D-1}{2}}}{\Gamma \left (\frac{D-1}{2} \right )}\int_{0}^{\infty}
u^{D-2} f(u) du,
\end{equation}

\noindent one can express the vacuum energy density as
\begin{equation}
E^{D}=\frac{2^{[(D+1)/2]-(D+1)}\Gamma\left ( -\frac D 2 \right
)}{\pi ^{\frac D 2}a\sqrt{\gamma}}\sum_{n=-\infty}^{\infty}\left
[\left (\frac{2 \pi (n+\frac 12)}{a \sqrt{\gamma}} \right
)^2+m_0^2\right ]^{\frac D2}
\end{equation}

It is seen from Eq. (15) that the expression for the vacuum energy
in the case of helix boundary conditions can be obtained from the
corresponding expression in the case of standard boundary condition
$\psi (t,x^1+a,x^2,x^T)=-\psi (t,x^1,x^2,x^T)$ by making the change
$a \rightarrow a\sqrt{\gamma }=\sqrt{a^2+h^2}$. The topological
fermionic Casimir effect in toroidally compactified space-times has
been recently investigated in Ref. \cite{Bellucci} for non-helix
boundary conditions including general phases. In the limiting case
$h=0$, our result of Eq. (18) is a special case of general formulas
from Ref.\cite{Bellucci}.

\section{The Case of massless field}

For a massless Dirac field, that is , in the case of $m_0=0$, the
energy density in Eq.(18) is reduced to
\begin{equation}\label{fienergy}
E^{D}_0=\frac {2^{[(D+1)/2]} \pi^{\frac D 2}}{a^{D+1}\gamma^{\frac
{D+1}{2}}}\Gamma \left (-\frac D 2 \right )\zeta(-D,\frac 12).
\end{equation}

\noindent where $\zeta(-D,\frac 12)$ is the Hurwitz-Riemann $\zeta$
function. Using the relation

\begin{equation}
\zeta(s,\frac 12)=(2^s-1)\zeta(s),
\end{equation}

\noindent and the reflection formula
\begin{equation}\label{rel}
    \Gamma\left(\frac{s}{2}\right)\zeta(s) = \pi^{s-\frac{1}{2}} \Gamma\left(\frac{1-s}{2}\right)\zeta(1-s)\,.
\end{equation}

 \noindent the energy density can be regularized to be

\begin{equation}\label{fienergy}
E^{D}_{R,0}=2^{[(D+1)/2]} \left (2^{-D}-1\right )\frac {\Gamma \left
(\frac {D+1} 2 \right )\zeta(D+1)}{\pi^{\frac {D+1}
2}(a^2+h^2)^{\frac{D+1}2}}.
\end{equation}
The Casimir force on the $x$ direction is
\begin{eqnarray}
F_{a,0}&=&-\frac {\partial E_{R,0}^{D}}{\partial a} =2^{[(D+1)/2]} \left (2^{-D}-1\right )\nonumber\\
&\times&\frac {(D+1)\Gamma \left (\frac {D+1} 2\right
)\zeta(D+1)}{\pi^{\frac {D+1} 2}} \frac{a}{(a^2+h^2)^{\frac{D+3}2}}
\end{eqnarray}

\noindent It is obvious that the energy density is negative and the
force is attractive. Furthermore, the force has a maximum value
\begin{eqnarray}
F_{a,0}^{max}&=&2^{[(D+1)/2]} \left (2^{-D}-1\right )\frac
{(D+1)\Gamma \left (\frac {D+1} 2 \right )\zeta(D+1)}{\pi^{\frac
{D+1}
2}h^{D+2}}\nonumber\\
&\times&\sqrt{\frac{(D+2)^{D+2}}{(D+3)^{D+3}}}
\end{eqnarray}
at $a=\frac{h}{\sqrt{D+2}}$. The results for $F_{h,0}$ are similar
to those of $F_{a,0}$ because of the symmetry between $a$ and $h$.

Fig. 1 is the illustration of the behavior of the Casimir force on
$x$ direction in $D=3$ dimension. The curves correspond to
$h=0.9,1.0,1.1,1.2$ respectively. It is clearly seen that the
attractive Casimir force decreases with $h$ increasing and the
maximum value of the force $-\frac{35\sqrt{5}\pi^2}{1944h^5}$
appears at $a=\frac{h}{\sqrt{5}}$.

Fig. 2 is the illustration of the behavior of the Casimir force on
$x$ direction in different dimensions. The curves correspond to
$D=2,3,4,5$ respectively. We take $h=2.5$ in this figure. It is
clearly seen that the value of $a$ where the maximum value of the
force is achieved gets smaller with $D$ increasing.

\begin{figure}
\includegraphics[width=0.5\textwidth]{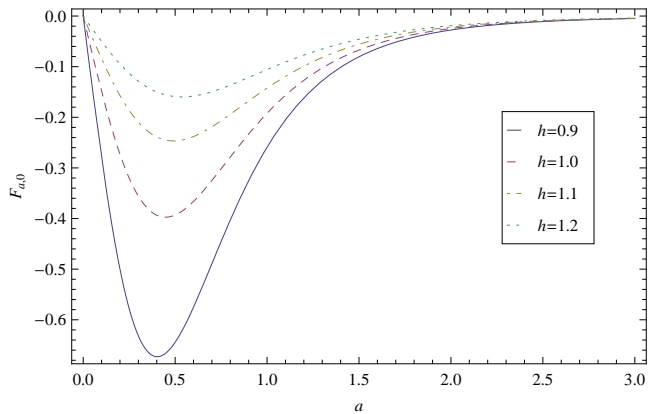}
\caption{The fermionic Casimir force on the $x$ direction
\textit{vs.} $a$ in $D=3$ dimension for different $h$. The Casimir
force decreases with $h$ increasing and the maximum value of the
force $-\frac{35\sqrt{5}\pi^2}{1944h^5}$ appears at
$a=\frac{h}{\sqrt{5}}$.} \label{fig:1}
\end{figure}

\begin{figure}
\includegraphics[width=0.5\textwidth]{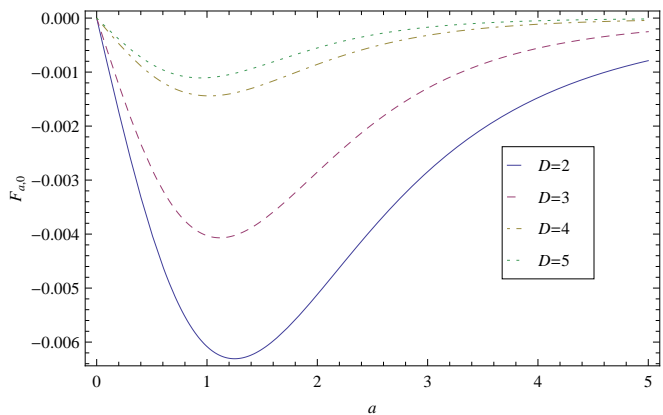}
\caption{The fermionic Casimir force on the $x$ direction
\textit{vs.} $a$ in different dimensions. Here we take $h=2.5$. It
is clearly seen that the value of $a$ where the maximum value of the
force is achieved gets smaller with $D$ increasing.}\label{fig:2}
\end{figure}

\section{The Case of massive field}

For a massive Dirac field, to regularize the series in Eq. (18) we
use the Chowla-Selberg formula directly \cite{Elizalde4}
\begin{eqnarray}
& &\sum_{n=-\infty}^{\infty}\left [\frac 1 2 a(n+c)^2+b\right ]^{-s}\nonumber\\
&=&
\frac{(2\pi)^{\frac 1 2}b^{\frac 1 2-s}}{\sqrt{a}}\frac{\Gamma\left(s-\frac 1 2\right)}{\Gamma(s)}
+\frac{2^{\frac s 2 +\frac 1 4+2}\pi^sb^{-\frac s 2+\frac 1
4}}{\sqrt{a}\Gamma(s)}\nonumber\\
&\times&\sum_{n=1}^{\infty}\cos(2\pi nc)\left(\frac
{n^2}{a}\right)^{\frac s 2-\frac 1 4}K_{\frac 1 2-s}\left(2\pi n\sqrt{\frac{2b}{a}}\right)
\end{eqnarray}
\noindent where $K_{\nu}(z)$ is the modified Bessel function. Note
that in the renormalization procedure, the vacuum energy in a flat
space-time with trivial topology should be renormalized to zero,
that is, in the expression for the renormalized vacuum energy the
term corresponding to the first term in the right hand side of Eq.
(25) should be ommited. Finally, the Casmir energy has the
expression as follows

\begin{eqnarray}
E_{R,m_0}^D&=&2^{[(D+3)/2]}\left
(\frac{m_0}{2\pi\sqrt{a^2+h^2}}\right )^{\frac{D+1}{2}}\nonumber\\
&\times&\sum_{n=1}^{\infty}\cos(\pi
n)n^{-\frac{D+1}{2}}K_{\frac{D+1}{2}}\left ( nm_0
\sqrt{a^2+h^2}\right)
\end{eqnarray}

\noindent For $\nu>0$ and $z\rightarrow 0$, the Bessel function has
the asymptotic expression $K_{\nu}(z)\rightarrow
\frac{2^{\nu-1}\Gamma(\nu)}{z^{\nu}}$, so it is not difficult to
find that when $m_0\rightarrow 0$, the Casimir energy recover the
result of the massless case.

 Using the relation
$K_{\nu}^{\prime}(z)=\frac{\nu}{z}K_{\nu}(z)-K_{\nu+1}(z)$ where
$K_{\nu}^{\prime}(z)=dK_{\nu}(z)/dz$, we have the Casimir force

\begin{eqnarray}
F_{a,m_0}&=&\frac{2^{[(D+3)/2]} m_0 a\left ( (m_0 a)^2+(m_0
h)^2\right )^{\frac{D+1}{4}}}{(2\pi)^{\frac{D+1}{2}}\left (
a^2+h^2\right )^{\frac {D+2}{2}}}\nonumber\\
&\times&\sum_{n=1}^{\infty}\cos(\pi
n)n^{-\frac{D-1}{2}}K_{\frac{D+3}{2}}\left ( n m_0
\sqrt{a^2+h^2}\right )\nonumber\\
\end{eqnarray}

We study numerically the behavior of the Casimir force on $x$
direction as a function of $a$ for different $h$ and $D$. We find
that the Casimir force is still attractive and it has a maximum
value similarly to massless case. Because the precise way the
Casimir force varies as the mass changes is worth studying, we give
the numerical results in Figs. 3 and 4 for the ratio of the Casimir
force in massive and massless cases varying with the mass for
different $h$ and $D$.

Fig.3 is the illustration of the ratio of the Casimir force in
massive case to that in massless case varying with the mass in $D=3$
dimension. The curves correspond to $a=1$ and $h=0.1,1,2,3$
respectively. Fig.4 is the illustration of the ratio of the Casimir
force in massive case to that in massless case varying with the mass
for different dimensions. The curves correspond to $a=1, h=0.1$ and
$D=2,3,4,5$ respectively. It is clearly seen from the two figures
that the Casimir force decreases with $m_0$ increasing, and it
approaches zero when $m_0$ tends to infinity. The plots also tell us
that for a given mass, the ratio decreases with $h$ increasing but
it increases with $D$ increasing.

\begin{figure}[h]
\includegraphics[width=0.5\textwidth]{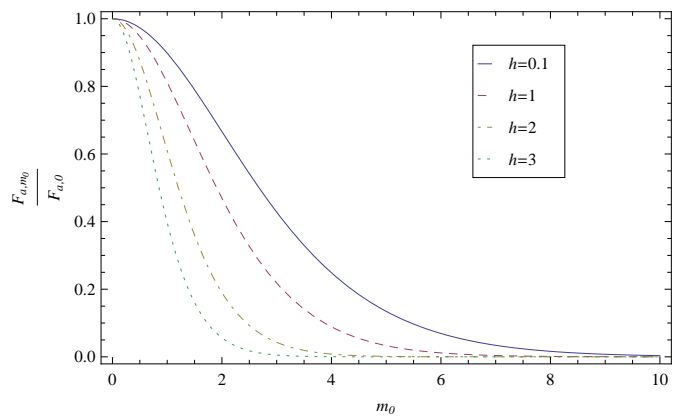}
\caption{The ratio of the Casimir force in massive case to that in
massless case varying with the mass for different $h$ in $D=3$
dimension. The curves correspond to $a=1$ and $h=0.1,1,2,3$
respectively.} \label{fig:3}
\end{figure}

\begin{figure}[h]
\includegraphics[width=0.5\textwidth]{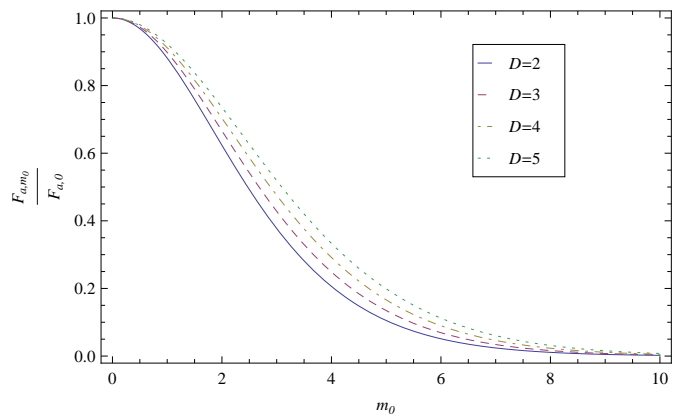}
\caption{The ratio of the Casimir force in massive case to that in
massless case varying with the mass for different dimensions. We
take $a=1$ and $h=0.1$. } \label{fig:4}
\end{figure}

\section{Conclusion }

The Casimir force arises not only in the presence of material
boundaries, but also in spaces with non-Euclidean topology and the
attractive or repulsive nature of the Casimir force may depend on
the topology of space-time. We have studied the Casimir effect of a
Dirac fermionic field under the helix topology. We saw that the
fermionic Casimir force is attractive under the helix boundary
condition.

Another interesting character is that the fermionic Casimir force
has a maximum value and the force decreases with $D$ increasing.
Equally interesting is that there is a $Z_2$ symmetry of
$a\leftrightarrow h$. Besides, when the mass of fermion tends to
zero, the Casimir force approaches the result of the force in
massless case and when the mass tends to infinity, the Casimir force
for a massive field goes to zero.

In recent years much attention has been paid to the possibility that
a universe could have non-trivial topology. As is known that the
Casimir effect can apply to the cosmology with extra dimensions,
 the effect of the quantum spring in higher dimensional cosmology is worth
  considering and we will study it in our further work.

\section*{Acknoledgments}

This work is supported by National Nature Science Foundation of
China under Grant No. 10671128 and No. 11047138, National Education
Foundation of China grant No. 2009312711004, Key Project of Chinese
Ministry of Education (No. 211059), Shanghai Natural Science
Foundation, China grant No. 10ZR1422000 and Innovation Program of
Shanghai Municipal Education Commission(11zz123).

\end{document}